\documentclass{amsart}
\usepackage{eurosym}
\usepackage{amsfonts}

\newtheorem{theorem}{Theorem}
\theoremstyle{plain}

\newtheorem{criterion}{Criterion}

\newtheorem{example}{Example}

\newtheorem{proposition}{Proposition}

\numberwithin{equation}{section}

\begin{document}
\title[ Probabilistic, statistical and algorithmic aspects of the similarity]{Probabilistic, statistical and algorithmic aspects of the similarity of
texts and application to Gospels comparison}
\author{Gane Samb Lo $^{**}$}
\author{Soumaila Dembele $^{*}$}
\email[G. S. LO]{ganesamblo@ganesamblo.net}
\email[S. Demebele]{soumidemlpot@gmail.com}
\address{$^{**}$ LSTA, Universit\'e Pierre et Marie Curie, France and LERSTAD, Universit\'e Gaston Berger
de Saint-Louis, SENEGAL\\
gane-samb.lo@ugb.edu.sn, ganesamblo@ganesamblo.net}
\address{$^{*}$ LERSTAD, Universit\'e Gaston Berger de Saint-Louis, SENEGAL and Universit\'e des Sciences de Gestion de Bamako, Mali}

\subjclass[2010]{62-07; 76M55}
\keywords{Similarity , Web mining, Jaccard similarity, RU algorithm, minhashing, Data mining, shingling, bible's Gospels, Glivenko-Cantelli, expected similarity, statistical estimation}

\begin{abstract}
The fundamental problem of similarity studies, in the frame of data-mining,
is to examine and detect similar items in articles, papers, books, with huge
sizes. In this paper, we are interested in the probabilistic, and the statistical
and the algorithmic aspects in studies of texts. We will be using the 
approach of $k$\textit{-shinglings}, a $k$\textit{-shingling} being defined
as a sequence of $k$ consecutive characters that are extracted from a text ($k\geq 1$
). The main stake in this field is to find accurate and quick algorithms to compute the
similarity in short times. This will be achieved in using approximation
methods. The first approximation method is statistical and, is based on the theorem
of Glivenko-Cantelli. The second is the banding technique. And the third concerns
a modification of the algorithm proposed by Rajaraman and al (%
\cite{AnandJeffrey}), denoted here as (RUM). The Jaccard index is the one used
in this paper. We finally illustrate these results of the paper on the four
Gospels. The results are very conclusive.
\end{abstract}

\maketitle




\section{Introduction}

\noindent In the modern context of open publication, in Internet in
particular, similarity studies between classes of objects become crucial.
For example, such studies can detect plagiarism of books, of articles, and
of other works. Also they may reveal themselves as decision and management
tools. Another illustration of the importance of such a knowledge concerns
commercial firms. They may be interested in similarity patterns between
clients from different sites or between clients who buy different articles.
In the same order of ideas, movies renting companies may try to know the
extent of similarity between clients subscribing for violence films and
those renting action films for example.\newline

\noindent As a probability concept, the notion of similarity is quite
simple. However in the context of Internet the data may be huge. So that the
main stake is the quick determination of some similarity index. The shorter
the time of computation, the better the case. So similarity studies should
rely on powerful algorithms that may give clear indications on similarities
in seconds. The contextualization of the similarity, and forming the sets to
be compared, and the similarity computations may take particular forms
according to the domains of application.\newline

\noindent In this paper, we will be focusing on similarity of texts. This
leads us to consider the approach of \textit{shinglings}, that we will define
in Section 2.\newline

\noindent The reader is referred to Rajaraman and \textit{al.} (\cite%
{AnandJeffrey}) for a general introduction to similarity studies. In their
book, they provide methods of determination of approximated indices of
similarity. Also, they propose an algorithm that we denote as RU (for
Rajaraman and Ullman). However this algorithm has not been yet investigated
in the context of probability theory, up to our knowledge. Furthermore, an
evaluation of the performances of such algorithms on usual texts may be of
relevance to justify such methods.\newline

\noindent First, we want to review these methods in a coherent probabilistic
and statistical setting allowing to reach - later - all the aspects of
similarity in this field. Then we will describe the RU algorithm in details.
We will point out its redundant sides, from which a modified algorithm -
denoted RUM (for RU modified) - will be proposed.\newline

\noindent To evaluate the studied techniques, the four Gospels will be used
with the ends of study of similarity. The techniques will be compared in
terms of speed, request of time, request of computer science resources, and
request of precision.\newline

\noindent The obtained results constitute a plea for improving these
techniques when dealing with larger sizes. \newline

\noindent Regarding Gospels study, our results seem to be conclusive, that is
the fourth canonical Gospels are significantly similar.\newline

\noindent This paper is organized as follows. In the next section, we define
the similarity of Jaccard and its metric and probabilistic approaches.
Section 3 is concerned with the similarity of texts. In Section 4, we
discuss about computation stakes of similarity. In Section 5, we present
different methods to estimate the similarity index. Finally in Section 7, we
deal with applications of the described methods to the similarity between
the four Gospels. We conclude the paper by giving some perspectives.\newline

\section{Similarity of sets}

\subsection{Definition}

\noindent Let A and B be two sets. The Jaccard similarity of sets A and B,
denoted $sim(A,B),$ is the ratio of the size of the intersection of A and B
to the size of the union of A and B:

\begin{equation}
sim(A,B)=\frac{\#(A\cap B)}{\#(A\cup B)}.  \label{sim1}
\end{equation}

\noindent It is easy to see that for two identical sets, the similarity is $%
100\%$ and for two totally disjoint sets, it is $0\%$.\newline

\subsection{Metric approach}

\noindent Let us consider a non-empty set $S$ and its power set $\mathcal{%
P(S)}$. Let us consider the application of dissimilarity:

\begin{equation*}
\forall \text{ }(S_{1},S_{2})\in \mathcal{P(S)}%
^{2},d(S_{1,}S_{2})=1-sim(S_{1,}S_{2}).
\end{equation*}

\noindent We have this simple result.\newline

\begin{proposition}
The mapping $d$ is metric.
\end{proposition}

\noindent \textbf{Proof}. Proving this simple result is not so obvious one might think.
Indeed, special techniques are required to demonstrate the triangle
inequality. This is done, for example, in (\cite{GowerLegendre}), page $15$.
Here, we just outline the other conditions for a metric :\newline

\begin{enumerate}
\item First, let us show that $\forall $ $(S_{1},S_{2})\in (\mathcal{P(S))}%
^{2},\ d(S_{1},S_{2})$ $\geq 0$.\newline
We have 
\begin{equation*}
\#(S_{1}\cap S_{2})\leq \#(S_{1}\cup S_{2}).
\end{equation*}%
Then 
\begin{equation*}
\frac{\#(S_{1}\cap S_{2})}{\#(S_{1}\cup S_{2})}\leq 1.
\end{equation*}%
Next 
\begin{equation*}
1-\frac{\#(S_{1}\cap S_{2})}{\#(S_{1}\cup S_{2})}\geq 0.
\end{equation*}%
Therefore 
\begin{equation*}
\ d(S_{1},S_{2})\geq 0.
\end{equation*}

\item Let us show that $d(S_{1,}S_{2})=0\Longrightarrow S_{1}=S_{2}$. From $(%
\ref{sim1})$, we get

\begin{equation*}
\#(S_{1}\cap S_{2})=\#(S_{1}\cup S_{2})\Longrightarrow S_{1}=S_{2}.
\end{equation*}

\item Let us remark that $d(S_{1,}S_{2})=d(S_{2},S_{1})$, since we have $%
S_{1}\cap S_{2}=S_{2}\cap S_{1}$, and $S_{1}\cup S_{2}=S_{2}\cup S_{1}$,
that is : the roles of $S_1$ and $S_2$ are symmetrical in what precedes.
\end{enumerate}

\bigskip

\noindent So, studying of the similarity is equivalent to studying the
distance of dissimilarity $d$ between two sets.

\subsection{Probabilistic approach}

\label{ss23}

\noindent Let us give a probabilistic approach of the similarity. For that,
let us introduce the notion of the representation matrix. Let $n$ be the
size of the introduced set above.\newline

\noindent Let us consider $p$ subsets of $S$: $S_{1},...,S_{p}$. The
representation matrix of $S_{1},...,S_{p}$ consists in this:

\begin{itemize}
\item We form a rectangular array of $p+1$ columns.

\item We put $S,S_{1},...,S_{p}$ in the first row.

\item We put in the column of $S$ all the elements of $S$, that we might
write from $1$ to $n$ in an arbitrary order.

\item In the column of each $S_{i},$ we will put $1$ or $0$ on the row $i$
depending on whether the $i^{th}$ element of $S$ is in $S_{i\text{ }}$or
not. We then can see that for $h\neq k$, $(S_{h}\cup S_{k})$ is the number
of rows \ for which one of the columns of $S_{h}$ or $S_{k}$ has $1$ on them
and $(S_{h}\cap S_{k})$ is the number of rows \ for which the two columns of 
$S_{h}$ and $S_{k}$ have $1$ on them.
\end{itemize}

\bigskip

\noindent The illustration of the matrix representation is as follows : 
\newline

\begin{center}
\begin{tabular}{|l|l|l|l|l|l|l|l|l|}
\hline
Element & $S_{1}$ & $S_{2}$ & ... & $S_{h}$ & ... & $S_{k}$ & ... & $S_{p}$
\\ \hline
$1$ & $1$ & $0$ & $...$ & $0$ & $...$ & $1$ & $...$ & $1$ \\ \hline
$2$ & $0$ & $0$ & $...$ & $1$ & $...$ & $0$ & $...$ & $0$ \\ \hline
$...$ & $0$ & $...$ & $...$ & $...$ & $...$ & $...$ & $...$ & $...$ \\ \hline
$i$ & $1$ & $0$ & $...$ & $1$ & $...$ & $1$ & $...$ & $1$ \\ \hline
$...$ & $...$ & $...$ & $...$ & $...$ & $...$ & $...$ & $...$ & $...$ \\ 
\hline
$n$ & $0$ & $0$ & $...$ & $0$ & $...$ & $0$ & $...$ & $1$ \\ \hline
\end{tabular}

Table (2.1)\label{tab1}
\end{center}

\bigskip

\noindent Let us denote $(S_{ih})_{1\leq i\leq n}$ the column of $S_{h}$. We
obtain

\begin{equation*}
sim(S_{h},S_{k})=\frac{\#\{i,1\leq i\leq n,S_{ih}=S_{ik}=1\}}{\#\{i,1\leq
i\leq n,(S_{ih}+S_{ik}=1)+(S_{ih}=S_{ik}=1)\}}.
\end{equation*}

\bigskip

\noindent This formula can be written also in the following form :

\begin{equation*}
sim(S_{h},S_{k})=\frac{\#\{i,1\leq i\leq n,S_{ih}+S_{ik}=2\}}{\#\{i,1\leq
i\leq n,(S_{ih}+S_{ik}=1)+(S_{ih}+S_{ik}=2)\}}.
\end{equation*}

\bigskip

\noindent In the next theorem, we will establish that the similarity is a 
\textbf{conditional} probability.\newline

\begin{theorem}
\label{t01} Let us randomly pick a row $X$ among $n$ \ rows. Let $S_{X,h}$ be
the value of the row $X$ for a column $h$, $1\leq h\leq p$. Then the
similarity between two sets $S_{h\text{ }}$and $S_{k\text{ }}$ is the
probability of the event $(S_{X,h}=S_{X,k}=1)$ with respect to the event $%
(S_{X,h}\cup S_{X,k}\geq 1)$. i.e 
\begin{equation*}
sim(S_{k\text{ }},S_{h})=\mathbb{P}[(S_{X,h}=S_{X,k}=1)/(S_{X,h}+S_{X,k}\geq
1)].
\end{equation*}
\end{theorem}

\noindent \textbf{Proof.} We first observe that for the defined matrix below, the set of
rows can be split into three classes, based on the columns $S_{k\text{ }}$%
and $S_{h}$:\newline

\noindent 1. The rows $X$ such as we have $(1,1)$ on the two places for
columns $S_{k\text{ }}$ and $S_{h}$.\newline

\noindent 2. The rows $Y$ such as we have $(1,0)$ or $(0,1)$ on the two
places for columns $S_{k\text{ }}$ and $S_{h}$.\newline

\noindent 3. The rows $Z$ such as we have $(0,0)$ on the two places for
columns $S_{k\text{ }}$ and $S_{h}$.\newline

\noindent Let us show that $sim(S_{k\text{ }},S_{h})=\mathbb{P}%
[(S_{X,h}=S_{X,k}=1)/(S_{X,h}\cup S_{X,k}\geq 1)]$. \newline

\noindent Clearly, the similarity is the ratio of the number of rows $X$ to
the sum of the numbers of rows $X$ and the number of rows $Y$. The rows $Z$
are not involved in the similarity between $S_{h}$ and $S_{k}$. Thus

\begin{equation*}
sim(S_{k\text{ }},S_{h})=\frac{\#\{i,1\leq i\leq n,S_{Xh}=1,S_{Xk}=1\}}{%
\#\{i,1\leq i\leq n,(S_{Xh}+S_{Xk}=1)+(S_{Xh}=1,S_{Xk}=1)\}}.
\end{equation*}

\bigskip

\noindent Then, by dividing the numerator and the denominator by $n$, we
will have

\begin{equation*}
sim(S_{k\text{ }},S_{h})=\frac{\frac{\#\{i,1\leq i\leq n,S_{Xh}=1,S_{Xk}=1\}%
}{n}}{\frac{\#\{i,1\leq i\leq n,(S_{Xh}+S_{Xk}=1)+(S_{Xh}=1,S_{Xk}=1)\}}{n}}.
\end{equation*}

\noindent Hence we get the result

\begin{equation*}
sim(S_{k\text{ }},S_{h})=\mathbb{P}[(S_{X,h}=S_{X,k}=1)/(S_{X,h}\cup
S_{X,k}\geq 1)].
\end{equation*}

\bigskip

\noindent This theorem will be the foundation of statistical estimation of
the similarity as a probability.\newline

\noindent \textbf{Important remark.} When we consider the similarity of two subsets, say $S_h$ and $S_k$
and we use the global space as $S_h \cup S_k$, we may see that the
similarity is, indeed, a probability. But when we simultaneously study the
joint similarities of several subsets, say at least $S_h$, $S_k$ and $S_\ell$
with the global set $S_h \cup S_k \cup S_\ell$, the similarity between two
subsets is a \textbf{conditional} probability. Then, using the fact that the
similarity is a probability to prove the triangle inequality is not
justified, as claimed in \cite{AnandJeffrey}, page 76.

\subsection{Expected similarity}

\label{ss224}

\noindent Here we shall use the language of the urns. Suppose that we have a
reference set of size $n$ that we consider as an urn \textbf{U}. We pick at
random a subset $X$ of size $k$ and a subset $Y$ of size $m.$ If $m$ and $k$
have not the same value, the picking order of the first set does have an
impact on our results. We then proceed at the beginning by picking at random
the first subset, that will be picked all at once, next put it back in the
urn \textbf{U} (reference set). Then we pick the other subset. Let us ask
ourselves the question : what is the expected value of the similarity of
Jaccard?\newline

\noindent The answer at this question allows us later to appreciate the
degree of similarity between the texts. We have the following result :%
\newline

\bigskip

\begin{proposition}
Let $U$ be a set of size $n$. Let us randomly pick two subsets $X$ and $Y$ of $%
U$, of respective sizes $m$ and $k$ according to the scheme described above.
We have 
\begin{equation}
\mathbb{P}(Card(X\cap Y)=j)=%
\begin{tabular}{lll}
$\frac{1}{2}\left\{ \frac{C_{m}^{j}\text{ \ \ }C_{n-m}^{k-j}}{C_{n}^{k}\text{
\ \ \ \ }C_{n}^{m}}\text{ \ \ }+\text{ }\frac{C_{k}^{j}\text{ \ \ }%
C_{n-k}^{m-j}}{C_{n}^{k}\text{ \ \ \ \ }C_{n}^{m}}\right\} $ & if & $0\leq
j\leq \min (k,m)$ \\ 
$0$ & otherwise & 
\end{tabular}%
.  \label{sim01}
\end{equation}%
Further 
\begin{equation}
\mathbb{E}(sim(X,\text{ \ \ }Y))=\sum_{j=0}^{\min (k,m)}\frac{j}{2(m+k-j)}%
\left\{ \frac{C_{m}^{j}\text{ \ \ }C_{n-m}^{k-j}}{C_{n}^{k}\text{ \ \ \ \ }%
C_{n}^{m}}\text{ \ \ }+\text{ }\frac{C_{k}^{j}\text{ \ \ }C_{n-k}^{m-j}}{%
C_{n}^{k}\text{ \ \ \ \ }C_{n}^{m}}\right\}  \label{sim02}
\end{equation}
\end{proposition}

\noindent \textbf{Proof}. Let us use the scheme described above. Let us first pick the set $X$. We
have $L=C_{n}^{k}$ possibilities. Let us denote the subsets that would take $%
X$ by $X_{1},...,X_{L}.$ The searched probability becomes%
\begin{equation*}
\mathbb{P}(Card(X\cap Y)=j)=\sum_{s=1}^{L}\mathbb{P}((Card(X\cap Y)=j)\cap
X_{s})=\sum_{s=1}^{L}\mathbb{P}((Card(X\cap Y)=j)/X_{s})\mathbb{P}(X_{s})
\end{equation*}%
Once $X_{s}$ is chosen and fixed, we get%
\begin{equation*}
\mathbb{P}((Card(X\cap Y)=j)/X_{s})=\frac{C_{m}^{j}\text{ \ \ }C_{n-m}^{k-j}%
}{C_{m}^{k}}\text{ }.
\end{equation*}%
Since $\mathbb{P}(X_{s})=1/C_{n}^{k}=1/L$, we conclude 
\begin{equation*}
\mathbb{P}(Card(X\cap Y)=j)=\sum_{s=1}^{L}\frac{C_{m}^{j}\text{ \ \ }%
C_{n-m}^{k-j}}{C_{m}^{k}}(1/L)=\frac{C_{m}^{j}\text{ \ \ }C_{n-m}^{k-j}}{%
C_{n}^{k}\text{ \ \ \ \ }C_{n}^{m}}.
\end{equation*}%
The result corresponding to picking up $Y$ first, is obtained by symmetry of
roles of $k$ and $n$. We then get (\ref{sim01}). The formula (\ref{sim02})
comes out immediately since%
\begin{equation}
sim(X,\text{\ }Y)=\frac{\#(X\cap Y)}{\#(X\cup Y)}=\frac{\#(X\cap Y)}{%
m+k-\#(X\cap Y)}.  \label{esp}
\end{equation}

\section{Similarity of texts:}

\label{s3}

\noindent The similarity is an automatic tool to anticipate the plagiarism,
abusive quotations, influences, etc. However the study of the similarity of
texts relies for instance on the words and not on the meanings.

\subsection{Forming of sets for comparison}

\label{ss31}

\noindent If we want to compare two texts $S_{1}$ and $S_{2},$ we must
transform them in \textit{shinglings } sets. For $k>0$, a $k$-\textbf{%
\textit{shingling}} is simply a word of $k$ letters. For finding the $k$-%
\textit{shinglings } of a string, we first consider the word of $k$
letters beginning with the first letter, the word of $k$ letters beginning
with the second letter, the word of $k$ letters beginning with the third,
etc.., until the word of $k$ letters finishing by the last letter of the
string. So, a string of $n$ letters is transformed into $(n-k+1)$ $k$-\textit{%
shinglings}.\newline

\noindent We observe a serious difficulty in the practice in using the
notion of similarity defined on \textbf{sets} of $k$-\textit{shinglings}.
Indeed, when we consider the $k$-\textit{shinglings} of a text, it is very
probable that some $k$-\textit{shinglings} will be repeated. Then the
collection of $k$-\textit{shinglings} cannot define a \textit{mathematical set}
(whose elements are supposed to be distinct).\newline

\noindent But fortunately, a $k$-\textit{shingling} is determined by its
value and its rank. Suppose that a text has a length $n$. We can denote the $%
k$-\textit{shinglings} by means of a vector $t$ of $n-k+1$ dimensions so
that $t(i)$ is the $i^{th}$ $k$-\textit{shingling}. The $k$-\textit{%
shinglings }set is defined by:

\begin{equation*}
\left\{ (i,t(i)),i=1,......,n-k+1\right\}
\end{equation*}

\noindent With this definition, the $k$-\textit{shinglings} are different
and do form a \textit{well-defined mathematical set}.

\subsection{Interpretation of \ the similarity of texts}

\label{ss32}

\noindent Does the similarity between two texts have necessarily another
explanation other than randomness? To answer to this question, let us remark
that in any language, a text is composed from an alphabet that is
formed by a finite and even small number of characters. A text in English is
a sequence of lowercase and uppercase letters of the alphabet, of numbers
and of some signs such as punctuations, apostrophes, etc. This set doesn't
exceed a hundred characters. \newline

\noindent Suppose that the computed similarity between the two\textit{\ }%
sets \textit{\ is } $p_{0}$. From what point can we reasonably consider that
there is a possible collision between the authors, either the two texts are
based on similar sources, or one author has used the materials of the other?
To answer this question, we have to know the part due to randomness. As a
matter of fact, any text is written from a \textit{\ }limited set of $k$-%
\textit{shinglings}. Then each $k$-\textit{shingling} is expected to occur
many times and hence contributes to rise the similarity. Let us consider a
set of size $n=m+\ell $ $k$-\textit{shinglings} containing those of the two
compared texts. If the two texts were randomly written, that is the same to
saying that they were written by machines subjected to randomness, the
expected similarity that we denote by $p_{R}$ would be given by (\ref{sim02}%
). So we can say that the two authors would have some kind relationship of
mutual influence or that plagiarism is suspected, if $p_{0}$ is
significantly greater than $\ p_{R}.$\newline

\noindent It is therefore important to have an idea of the value of $p_{R}$
for sizes of the order of those of studied texts. For example, with the
Bible texts that we study, the texts sizes go approximately from\ $50.000$
to $110.000.$ The values $p_{R}$ for these sizes turn round $30\%$. This
knowledge is important to interpret the results.

\subsection{ Implementation of the algorithm for computing the similarity of
texts}

\label{ss33}

\noindent Let $A$ and $B$ \ be two texts. Fixed $k\geq 1$ and let us
consider the two $k$-\textit{shinglings} sets\textit{\ }%
\begin{equation*}
((i,t_{A}(i)),i=1,......,n_{A}-k+1)
\end{equation*}

\noindent and $\ $

\begin{equation*}
((j,t_{B}(j)),j=1,......,n_{B}-k+1).
\end{equation*}
\newline

\noindent The determination of the similarity between the two texts is
achieved through comparing each $k$-\textit{shingling }of $A$ with all $k$-%
\textit{shinglings} of $B$. We will have two problems to solve.\newline

\noindent Suppose that a $k$-\textit{shingling} is represented many times in 
$B$. We have the risk that the same value of this $k$-\textit{shingling} in $%
A$ is used as many times when forming the intersection between of $k$-%
\textit{shinglings} sets. This would result in a disaster.\newline

\noindent To avoid that, we associate to each $k$-\textit{shingling} $%
(i,t_{A}(i))$ at most one $k$-\textit{shingling} $(j,t_{B}(j))$. Let us
use the wedding language by considering the $k$-\textit{shinglings } of $A$
as husbands, and the $k$-\textit{shinglings} of $B$ as wives and, then, the
association between a $k$-\textit{shingling} of $A$ to a $k$-\textit{%
shingling} of $B$ as a wedding.\textit{\ } Our principle says that a $k$-%
\textit{shingling} of $A$ can marry at most one $k$-\textit{shingling}
of $B$. In the same way, a $k$-\textit{shingling} of $B$ can be married at
most to one $k$-\textit{shingling} of $A.$ We are in a case of perfect
symmetry monogamy. How to put this in practice in a program? \newline

\noindent It suffices to introduce the sentinel variables that identify if a 
$k$-\textit{shingling} husband or a $k$-\textit{shingling } wife has a wife
or a husband at the moment of the comparison.\newline

\noindent Let us introduce the vectors%
\begin{equation*}
(test_{A}(i),i=1,.....,n_{A}-k+1)
\end{equation*}

\noindent and%
\begin{equation*}
(test_{B}(j),j=1,........,n_{B}-k+1).
\end{equation*}

\bigskip

\noindent We put $test_{A}(i)=1$ if $k$-\textit{shingling} has already a
wife, $test_{A}(i)=0$ otherwise. We define $test_{B}(j)$ in the same manner.
We apply the following algorithm:

\begin{itemize}
\item[1.] set $sim=0$;

\item[2.] Repeat for $i=1$ to $n_{A}-k+1$;

\begin{itemize}
\item[2a.] if $test_{A}(i)=1$ \ : nothing to do;

\item[2b.] else \ 

\begin{itemize}
\item[2b-1.] do for : \ $j=1$ to $n_{B}-k+1$;

\begin{itemize}
\item[2b-11.] if $t_{B}(j)=1$ \ : nothing to do;

\item[2b-12.] else compare \ $t_{A}(i)$ to $t_{B}(j)$;

\item[2b-13.] if equality holds, increment $sim$ and put $test_{A}(i)=1$, $%
test_{B}(j)=1 $;

\item[2b-14.] else go to the next $j$.
\end{itemize}
\end{itemize}
\end{itemize}

\item[3.] report the similarity $(sim/(n_{A}+n_{B} - sim))$
\end{itemize}

\section{Computation stakes}

\label{s4}

\noindent The search of similarity faces many challenges in the Web context
and at the local post of personal computer.\newline

\subsection{Limitation of the random access memory (RAM)}

\label{ss41}

\noindent When we want to compare two sources of texts, each leading to a
large number of \textit{shinglings}, say $n_{1}$ and $n_{2}$, using the
direct method will load in memory the vectors $t_{A}$, $t_{B}$, $test(A)$
and $test(B)$. When $n_{1}$ and $n_{2}$ are very large with respect to the
capacities of the machine, this approach becomes impossible. For example,
for the values of $n_{1}$ and $n_{2}$ in order of $98 000 000$, the
declaration of vectors of that order leads to an overflow in Microsoft VB6$%
^{R}$. \newline

\noindent We are tempted to appeal to another method, that directly uses
data from files. Here is how it works:

\begin{itemize}
\item[(1)] open the file of the text $A$;

\item[(2)] read a row of the file $A$;

\item[(3)] open the file $B$: read all these rows one by one and compare the 
$k-$\textit{shinglings} of the file $B$ with the $k-$\textit{shinglings} at
the current row of file $A$.

\item[(4)] close the file $B$;

\item[(5)] go to the next row of file $A$.
\end{itemize}

\bigskip

\noindent This method that we denote by the similarity by file does
practically not use the RAM of the computer. We are then facing to two
competing methods. Each of them has its qualities and its defects.

\subsubsection{The direct method:}

\label{sss41}

\noindent It loads the vectors of $k-$\textit{shinglings} in the RAM. It
leads to quick calculations. However we have the risk to stuck the machine
when the sizes of the files are huge.

\subsubsection{The method of similarity by file}

\label{sss42}

\noindent It spares the RAM of the machine and increases the computing
speed. However it leads to considerable times of computations since, for
example, the second file is opened as many time as the first contains rows.
We spare the RAM but we lose time.\newline

\noindent You have to notice that in the implementation of this method, we
always have to carry the incomplete ends of each row at the next row.

\begin{example}
Suppose that we compare the $5-$\textit{shinglings} of the first row of
\noindent $A$ and the first row of $B$. The last four letters of the row
cannot form a $5-$\textit{shinglings}. We have to use them by adding them at
the first place of the second row of $A.$ These additional ends are denoted
"boutavant1"'s in the procedures done in $(\ref{ss41})$, when we implement
the similarity by file method. We do the same thing for the rows of $B$ that
give the "boutavant2"'s.
\end{example}

\bigskip

\noindent For example, in the work on the Gospel versions, where the numbers
of $k$-\textit{shinglings} are of the order of one hundred thousands, the
method of similarity by file takes around thirty minutes and the direct
method requires more or less ten minutes. We reduce the time of computation
by three at the risk to block the RAM.\newline

\noindent All what precedes advocates using approximated methods for
computing similarity. Here, we are going to see three approaches but we
only apply two of them in the study of the Gospel texts.

\section{Approximated computation of Similarity}

\label{s5}

\subsection{Theorem of Glivenko-Cantelli}

\label{ss51}

\noindent Since the similarity is a conditional probability in according to
Theorem $\ref{t01}$, we can deduce a law of Glivenko-Cantelli in the
following way.

\begin{theorem}
Let $p$ be the similarity between two sets of total size $n$. Let us pick at
random two subsets of respective sizes $n_{1}$ and $n_{2}$ so that $%
n=n_{1}+n_{2}$ and let us consider the random similarity $p_{n}$ between
these two subsets. Then $p_{n}$ converges almost-surely to $p$ with a speed
of convergence in the order of $n^{-1/4}$ when $n_1$ and $n_1$ become very
large.
\end{theorem}

\bigskip

\noindent That is a direct consequence of the classical theorem of
Glivenko-Cantelli. It then yields a useful tool. For example, for the
similarity of Gospels for which the similarity is determined in more or less
ten minutes, the random choice of subsets of size around ten thousand $k$-%
\textit{shinglings} for each Gospel gives a computation time less than
one minute, with an accuracy of $90\%$. To avoid the instability due to one
random choice only, the average on ten random choices gives a better
approximated similarity in more or less one minute. We will widely come back
to this point in the applications.

\bigskip

\subsection{ Analysis of the Banding Technique}

\label{ss52}

\noindent The banding technique is a supplementary technique based on the
approximation of Theorem of Glivenko-Cantelli. Suppose that we divide the
representation matrix, in $b$ bands of $r$ rows. The similarity can be
computed first by considering the similarity between the different rows of one
band then, between some bounds only. We do not use this approach here.

\subsection{Algorithm of RU}

\label{ss53}

\noindent It is based on the notion of minhashing to reduce documents of
huge sizes into documents of small sizes called signatures. The computation
of the similarity is done on their compressed versions, i.e, on their
signatures. To better explain this notion, let us consider $p$ subsets of a
huge reference set. Let the matrix be defined as below :\newline

\begin{center}
\begin{tabular}{|l|l|l|l|l|}
\hline
Element & $S_{1}$ & $S_{2}$ & . & $S_{3}$ \\ \hline
1 & 1 & 0 & . & 0 \\ \hline
2 & 0 & 0 & . & 1 \\ \hline
. & 0 & . & . & . \\ \hline
i & 1 & 0 & . & 1 \\ \hline
. & . & . & . & . \\ \hline
n & 0 & 0 & . & 0 \\ \hline
\end{tabular}

Table (5.1)\label{tab2}
\end{center}

\bigskip

\noindent The similarity between two sets is directly got as soon as this
table is formed by using the formula $(\ref{sim1})$ in a quick way. But the
setting of this matrix takes time. This is serious drawback of the original
algorithm RU that we will precise soon. For the moment, suppose that the table
exists. On this basis, we are going to introduce the RU algorithm. By this
algorithm, we do three things. First, we consider an arbitrary permutation
of the rows. Then, we replace the column of the rows by a transformation
called \textit{minhashes} by means of a congruence function. Then, a new table is
formed to replace the original table. This new and shorter one, that we
describe immediately below, is called signature matrix.

\subsubsection{ Minhashing signature}

\label{ss54}

\noindent Suppose that the elements of $S$ are given in a certain order
denoted from $1$ to $n.$ Let us consider $p$ functions $h_{i}$ $(i=1,...,p)$
from $\{1,...,n\}$ in itself in the following form:

\bigskip

\begin{equation}
h_{i}(x)=a_{i}x+b_{i}\text{ mod }n,  \label{cong}
\end{equation}

\bigskip

\noindent where $a_{i}$ and $b_{i}$ are given integers. We modify this
function in the following way: $h_{i}(x)=n$ when the remainder of the
euclidian division is zero. We then can transform the matrix as follows :

\bigskip

\begin{center}
\begin{tabular}{|l|l|l|l|l|l|l|l|}
\hline
Element & $S_{1}$ & $S_{2}$ & ... & $S_{m}$ & $h_{1}$ & $...$ & $h_{p}$ \\ 
\hline
1 & 1 & 0 & ... & 0 & $h_{1}(1)$ & ... & $h_{p}(1)$ \\ \hline
2 & 0 & 0 & ... & 1 & $h_{1}(2)$ & ... & $h_{p}(2)$ \\ \hline
. & 0 & . & ... & . & . & ... & . \\ \hline
i & 1 & 0 & ... & 1 & $h_{1}(i)$ & ... & $h_{p}(i)$ \\ \hline
. & . & . & ... & . & . & . & . \\ \hline
n & 0 & 0 & ... & 0 & $h_{1}(n)$ & ... & $h_{p}(n)$ \\ \hline
\end{tabular}

Table (5.2) \label{tab3}
\end{center}

\noindent The RU algorithm replaces this matrix by another smaller one
called \textbf{minhashing} signature, that is :

\bigskip

\begin{equation*}
\begin{tabular}{|l|l|l|l|l|}
\hline
hashing & $S_{1}$ & $S_{2}$ & $...$ & $S_{m}$ \\ \hline
$h_{1}$ & $c_{11}$ & $c_{12}$ & $...$ & $c_{1m}$ \\ \hline
$h_{2}$ & $c_{21}$ & $c_{22}$ & $...$ & $c_{2m}$ \\ \hline
... & $...$ & $.$ & $...$ & $.$ \\ \hline
$h_{p}$ & $c_{p1}$ & $c_{p3}$ & $...$ & $c_{pm}$ \\ \hline
\end{tabular}%
\end{equation*}

\begin{center}
\bigskip \qquad \qquad Table $(5.3)$ \label{tab4}
\end{center}

\noindent To fill the table above, Rajaraman and al.(\cite{AnandJeffrey}),
page 65, propose the algorithm below:

\bigskip

\textbf{Algorithm of filling of the columns }$S_{j}:$

\begin{itemize}
\item[1.] Set all the $c_{rj}$ equal to $\infty .$

\item[2.] For each column $S_{j},$proceed like this

\begin{itemize}
\item[2-a.] for each element $i,$ from $1$ to $n$, compute $%
h_{1}(i),h_{2}(i),........,h_{p}(i)$.

\item[2-b.] if $i$ is not in $S_{j},$ then do nothing and go to $i+1$

\item[2-c.] if $i$ is in $S_{j},$ replace all the rows $(c_{rj})_{1\leq
r\leq p} $ by the minimum: min($c_{rj},h_{r}(i)).$

\item[2-d.] go to $i+1$
\end{itemize}

\item[3.] go to $j+1$

\item[4.] end.
\end{itemize}

\bigskip

\noindent At the end of the procedure, each column will contain only
integers between $1$ and $n.$ The computed \ similarity on this compressed
table between $S_{i}$ and $S_{j}$, denoted simRU$(S_{1},S_{2}),$ will be
called approximated similarity RU. It is supposed to give an accurate
approximation of the similarity.

\bigskip

\noindent However we can simplify this algorithm in a very simple way \ by
saying this.

\begin{criterion}
\label{crit1} \textbf{The transpose of the column }$(c_{rj})_{1\leq r\leq
p},  $\textbf{is the minimum of rows, when carried out coordinate by
coordinate, }$(h_{1}(i),...,h_{p}(i)),$\ \textbf{when }$i$ \textbf{\ covers
the elements of }$S_{j}.$
\end{criterion}

\noindent This simple remark allows to set up programs in a much easier way.%
\newline

\subsubsection{Algorithm of RU modified (RUM)}

\label{ss55}

\noindent It is clear that by forming the matrix of the table (5.1), the
similarity is automatically computed. Indeed, when we consider the columns $%
S_{i}$ and $S_{j}$, we immediately see that the number of rows containing
the unit number (1) on these two columns is the size of the intersection.
Then the Jaccard similarity is already found and any further step is
useless. The RU algorithm, on this basis, is not useful. Instead, forming
this matrix is exactly applying the full method that requires comparison of
each couple of \textit{shinglings} of the two sets. This operation takes
about thirty minutes for set of sizes one hundred thousands, for example.
Based on this remark, we propose a modification for the implementation of
the RU algorithm in that following way. Let us consider two sets $S_{1}$ and 
$S_{2}$ with respective sizes $n_{1}$ and $n_{2}$ to be compared. We proceed
like that: \newline

\begin{itemize}
\item[1.] Form one set $S$ by putting the elements of $S_{1}$ and then the
elements of $S_{2}$ with the double elements. Let $n=n_{1}+n_{2}$.

\item[2.] Apply the RU algorithm at this collection by using Criterion \ref%
{crit1}
\end{itemize}

\bigskip

\noindent We do not seek to find the intersections. Elements of the
intersection are counted twice here. But it is clear that we still have a zero
similarity index if the two sets $S_{1}$ and $S_{2}$ are disjoint, and a $%
100\%$ index if the sets are identical.\newline

\noindent The question is : how well the estimations of the similarity using RU or RUM
algorithm are good approximations of the true similarity index? We give
in these paper an empirical response based on the Gospels comparison but
showing that the RUM approximation of the similarity of good while
performing only in a few seconds in place of thirty minutes (1.800 seconds)!%
\newline

\noindent The exact distribution of the RUM index is to be found depending
on the laws of the stochastic laws of the coefficients $a_i$ and $b_i$ in (%
\ref{cong}) in a coming paper.\newline

\bigskip

\section{The applications of the similarity of the Bible texts}

\label{s6}

\subsection{Textual context of the Gospel}

\label{ss61}

\textbf{\ Four versions of the Gospels}

\noindent Here, we are going to resume a few important points for the
backgrounds of our Gospels analysis. In all this subsection, we refer to 
\cite{EVan}.\newline

\noindent The Gospels (of Latin that means good news) are texts that relate
the life and the teaching of Jesus of Nazareth, called Jesus Christ. Four
Gospels were accepted as canonical by the churches: the Gospel according to
Matthew, Mark, Luke and John. The other unaccepted Gospels are qualified
apocryphal ones. Numerous Gospels have been written in the first century in
our era. Before to be consigned as written, the message of Christ was
verbally transmitted. From tale stories, many texts were composed, among
which the four Gospels that were retained in the Biblical canon. The
canonical Gospels are anonymous. They were traditionally attributed to
disciples of Jesus Christ. The Gospel according to Matthew and the Gospel
according to John would have been from direct witnesses of the preaching of
Jesus. Those of Mark and Luke are related to close disciples.

\bigskip

\noindent The first Gospel is the one attributed to Mark. It would have been
written in about 70 years AD. In about 80 - 85, follows the Gospel according
to Luke. The Gospel according to Matthew is dated between 80 and 90, and to
finish, the one of John is dated in between 80 and 110. However, these
uncertain dates vary according to the authors that propose chronologies of
the evangelical texts. The original Gospels were written in Greek.

\bigskip

\noindent The Gospel according to Matthew, Mark and Luke are called
Synoptic. They tell the tale of Jesus in a relatively similar way. The
Gospel according to John are written using another way of taling Jesus' life
and mission (christology) qualified as Johannist. The first set of Gospel
that has been written seems to be Mark's one. According to some researchers,
the common parts between Matthew and Luke Gospels may depend on a more older
text that was lost. This text is referred as the Q source.\newline

\noindent The source Q or Document Q or simply Q (The letter is from the
German word QUELLE, meaning source) is a hypothetical source, of whom some
exegetes think it would be at the origin of common elements of Gospels of
Matthew and Luke. Those elements are absent in Mark. It would be a
collection of words of Jesus of Nazareth that some biblists attempted to
reconstitute. This source is thought to date around of 50 AD.\newline

\noindent The Gospels of Matthew and Luke are traditionally influenced by
Mark's Gospel and the Old Testament. But though separately written, they
have in common numerous extracts that don't come from the two first cited
sources. This is why the biblists of XIX$^{\text{e}}$ century generally
think that these facts suggest the existence of a second common source,
called "document Q". Since the end of XIX$^{\text{e}}$ century, Logia (i.e
the speech in Greek) seems to have been an essentially collection of
speeches of Jesus. With the hypothesis of the priority of the Gospel of
Mark, the hypothesis of the existence of the document Q is part of what the
biblists call the hypothesis of two sources.

\bigskip

\noindent This hypothesis of two sources is the most general solution that
is accepted for the synoptic problem, that concerns the literary influences
between the three canonical Gospels ( Mark, Matthew, Luke), called Synoptic
Gospels. These influences are sensitive by the similarities in the choice of
words and the order of these words in the statement. The "\textbf{Synoptic
problem}" wonders about the origin and the nature of these relationships.
From the hypothesis of two sources, not only Matthew and Luke learned all
both on the Gospel according to Mark, independently one to other; but as we
detect similarities between the Gospels of Matthew and Luke, that we cannot
find in the Gospel of Mark, we have to suppose the existence of a second
source.

\bigskip

\noindent \textbf{Synoptic Gospels}\newline

\noindent The Gospels of Matthew, Mark, and Luke are considered synoptic
Gospels on the basis of many similarities between them that are not shared
by the Gospel of John. Synoptic means here that they can be seen or read
together, indicating the many parallels that exist among the three.

\bigskip

\noindent The Gospel of John, on the contrary has been recognized, for a
long time as distinct of first three Gospels so much by the originality of
its themes, of its content, of the interval of time that it recovers, and of
its narrative order and the style. Cl\'{e}ment of Alexandria summarized the
single character of the Gospel of John by saying : \textit{John came last,
and was conscious that the terrestrial facts had been already exposed in the
first Gospel. He composed a spiritual Gospel.}

\bigskip

\noindent Indeed, the fourth Gospels, the Gospel of John, presents a very
different picture of Jesus and his ministry from the synoptics. In
differentiating history from invention, some historians interpret the Gospel
accounts skeptically but generally regard the synoptic Gospels as including
significant amounts of historically reliable information about Jesus. The
common parts of the Gospels of Matthew and of Luke depend on an antiquarian
document but lost called source Q according to some researchers.

\bigskip

\noindent The synoptic Gospels effectively have many parallels between them:
thus around $80\%$ \ of verses of Mark may be found in Matthew and Luke
Gospels. As the content is in three Gospels, one talks about of Triple
tradition. The passages of the Triple Tradition are essentially narrations
but we can find in it some speeches of Christ.

\bigskip

\noindent But otherwise, we also find numerous identical passages between
Matthew and Luke, but absent in the Gospel of Mark. Almost $25\%$ of verses
of the Gospel according to Matthew find an echo from Luke (but not from
Mark). The common passages between Matthew and Luke are mentioned as the
Double Tradition.

\bigskip

\noindent The four Gospels constitute the principle documentary concerning
the life and the teaching \ of Christ. Each of them uses a particular
perspective. But all of them use the same general scheme and convey the same
philosophy. We stop here. For further details see \cite{EVan}. We will
attempt to explain the results in our own analysis of similarity below.

\subsection{The general setting}

\label{ss62}

\noindent All the computations  were done in the environment of VB6$^{R}$. Once the four texts are chosen,
we follow these steps. We first proceed to the editing files by dropping the
words of less than three letters. Then we proceed to the computations of the
similarity between the different Gospels.\newline

\noindent Next, we find for each gospel, the number of the rows of files as
well as the number of letters.\newline

\noindent Here is the first table for number of the rows, before and after
editing.\newline

\begin{center}
\begin{tabular}{|l|l|l|l|l|}
\hline
& John & Luke & Mark & Matthew \\ \hline
Numbers of the rows & 2534 & 3442 & 628 & 1319 \\ \hline
Numbers of letters before editing & 96269 & 129548 & 76543 & 149747 \\ \hline
Numbers of letters after editing & 69316 & 94766 & 55555 & 108722 \\ \hline
\end{tabular}

Table $(6.1)$ \label{tab5}
\end{center}

\noindent Now we are going to report the common numbers of $k$-\textit{%
shinglings} with $k=3$ between the different Gospels and then compute the
similarity between each couple of Gospels by the two exact methods.\newline

\noindent The results are in Tables 6.2 and 6.3.\newline

\noindent \textbf{Table : Case of computation of the similarity by the
direct method between the different Gospels}\newline

\begin{center}
\begin{tabular}{|l|l|l|l|}
\hline
& Luke & Mark & Matthew \\ \hline
John & Sim= 57,62 \% & Sim= 57,53 \% & Sim= 51 \% \\ 
& kc= 59981 & kc= 45600 & kc= 60134 \\ 
& time= 755 s & time= 816 s & time= 510 s \\ \hline
Luke &  & Sim=54,12 \% & Sim=69,55 \% \\ 
&  & kc= 52782 & kc= 83468 \\ 
&  & time= 640 s & time= 1430 s \\ \hline
Mark &  &  & Sim=48,74 \% \\ 
&  &  & kc= 53827 \\ 
&  &  & time= 508 s \\ \hline
\end{tabular}

Table $(6.2)$ \label{tab6}
\end{center}

\noindent \textbf{Table : Case of computation of the similarity by the
method by file between the different Gospels}\newline

\begin{center}
\begin{tabular}{|l|l|l|l|}
\hline
& Luke & Mark & Matthew \\ \hline
John & Sim= 57,62 \% & Sim= 57,53 \% & Sim= 51 \% \\ 
& kc= 59981 & kc= 45600 & kc= 60134 \\ 
& time= 2312 s & time= 1376 s & time= 3021 s \\ \hline
Luke &  & Sim=54,12 \% & Sim=69,55 \% \\ 
&  & kc= 52782 & kc= 83468 \\ 
&  & time= 3080 s & time= 1457 s \\ \hline
Mark &  &  & Sim=48,74 \% \\ 
&  &  & kc= 53827 \\ 
&  &  & time= 1552 s \\ \hline
\end{tabular}

Table $(6.3)$ \label{tab7}
\end{center}

\bigskip

\noindent \textbf{\ Approximated similarity}\newline

\noindent In this part, the computation of the similarity will be done by
the direct method. Let us pick randomly 10000 $k$-\textit{shinglings} from
first file and 10000 $k$-\textit{shinglings } from the second file. We
remark that the time of computation of the similarity turns around 20
seconds. We get approximated values of similarities between the Gospels. Let
us use the two methods of computation through a double approximation of the
similarity i.e, approximation using the theorem of Glivenko-Cantelli and of
the RUM algorithm. The two results are given in the two tables as follows:%
\newline

\noindent \textbf{Table : Case of computation of the approached similarity
by the theorem of Glivenko-Cantelli between the different Gospels}\newline

\begin{center}
\begin{tabular}{|l|l|l|l|}
\hline
& Luke & Mark & Matthew \\ \hline
John & Sim= 47,50 \% & Sim= 46,46 \% & Sim= 46,04 \% \\ 
& time= 20 s & time= 34 s & time= 29 s \\ \hline
Luke &  & Sim=50,79 \% & Sim=50,26 \% \\ 
&  & time= 19 s & time= 22 s \\ \hline
Mark &  &  & Sim=52,28 \% \\ 
&  &  & time= 27 s \\ \hline
\end{tabular}

Table $(6.4)$ \label{tab8}
\end{center}

\bigskip

\noindent \textbf{Table : Case of computation of the approximated similarity
by the RUM algorithm between the different Gospels }\newline

\noindent This approach is simply extraordinary since we may use a very low
number of hash functions and get good approximations. To guarantee the
stability of the results, we report the average results got for BB=50
repetitions of the experience and the standard deviation of such a sequence
of results.\newline

\noindent Case for pp=5 and BB=50.\newline

\begin{center}
\begin{tabular}{|l|l|l|l|}
\hline
& Luke & Mark & Matthew \\ \hline
John & Sim= 60 \% & Sim= 58 \% & Sim= 59,2 \% \\ 
& Ecart= 20,76 & Ecart= 22,1 & Ecart= 20,38 s \\ \hline
& time= 26 s & time= 25 s & time= 22 s \\ \hline
Luke &  & Sim= 56,4 \% & Sim= 63,2 \% \\ 
&  & Ecart= 19,87 & Ecart= 22,03 \\ \hline
&  & time= 22 s & time= 22 s \\ \hline
Mark &  &  & Sim= 59,2 \% \\ 
&  &  & Ecart= 22,38 \\ \hline
&  &  & time= 27 s \\ \hline
\end{tabular}

Table $(6.5)$ \label{tab9}
\end{center}

\bigskip

\noindent \textbf{Table: Case of computation of the approximated similarity
by the RUM algorithm \ between the different Gospels }\newline

\noindent Case for pp=10 and BB=50.\newline

\begin{center}
\begin{tabular}{|l|l|l|l|}
\hline
& Luke & Mark & Matthew \\ \hline
John & Sim= 62 \% & Sim= 61,6 \% & Sim= 63,6 \% \\ 
& Ecart= 16,68 & Ecart= 15,91 & Ecart= 14,52 s \\ \hline
& time= 26 s & time= 25 s & time= 22 s \\ \hline
Luke &  & Sim= 62 \% & Sim= 58 \% \\ 
&  & Ecart= 14,56 & Ecart= 13,41 \\ \hline
&  & time= 31 s & time= 27 s \\ \hline
Mark &  &  & Sim= 63,8 \% \\ 
&  &  & Ecart= 14,54 \\ \hline
&  &  & time= 29 s \\ \hline
\end{tabular}

Table $(6.6)$ \label{tab10}
\end{center}

\bigskip

\noindent \textbf{Table: Case of computation of the approximated similarity
by the RUM algorithm \ between the different Gospels }\newline

\noindent Case for pp=15 and BB=50.\newline

\begin{center}
\begin{tabular}{|l|l|l|l|}
\hline
& Luke & Mark & Matthew \\ \hline
John & Sim= 57 \% & Sim= 59,33 \% & Sim= 58,26 \% \\ 
& Ecart= 13,45 & Ecart= 11,33 & Ecart= 13,84 s \\ \hline
& time= 28 s & time= 28 s & time= 30 s \\ \hline
Luke &  & Sim= 60,13 \% & Sim= 58,26 \% \\ 
&  & Ecart= 13,23 & Ecart= 13,51 \\ \hline
&  & time= 30 s & time= 28 s \\ \hline
Mark &  &  & Sim= 57,2 \% \\ 
&  &  & Ecart= 14,17 \\ \hline
&  &  & time= 29 s \\ \hline
\end{tabular}

Table $(6.7)$ \label{tab11}
\end{center}

\bigskip

\noindent \textbf{Table: Case of computation of the approximated similarity
by the RUM algorithm \ between the different Gospels }\newline

\noindent Case for pp= 20 and BB=50.\newline

\begin{center}
\begin{tabular}{|l|l|l|l|}
\hline
& Luke & Mark & Matthew \\ \hline
John & Sim= 57,6 \% & Sim= 60,8 \% & Sim= 62,9 \% \\ 
& Ecart= 10,63 & Ecart= 10,11 & Ecart= 10,63 s \\ \hline
& time= 32 s & time= 31 s & time= 31 s \\ \hline
Luk &  & Sim= 57,9 \% & Sim= 63,6 \% \\ 
&  & Ecart= 9,59 & Ecart= 9,22 \\ \hline
&  & time= 32 s & time= 31 s \\ \hline
Mark &  &  & Sim= 60,7 \% \\ 
&  &  & Ecart= 8,94 \\ \hline
&  &  & time= 31 s \\ \hline
\end{tabular}

Table $(6.8)$ \label{tab12}
\end{center}

\subsection{ Analysis of results}

\label{ss63}

\subsubsection{Evaluation of algorithms}

\label{sss631}

\bigskip

\noindent \textbf{\ Algorithm on the similarity by the direct method}\newline

\noindent In this algorithm, we first form the $k$-\textit{shinglings} sets
for each text. Then we compute the similarity between them. \newline

\noindent We remark that the time of the determination of the similarity
between the different Gospels turns around ten minutes. The different
similarity amounts are around $50\%.$ \newline

\noindent \textbf{Algorithm on the similarity by the method by file}\newline

\noindent Here we remark that the times of the determination are much
greater than those in the case of the similarity by the direct method. The
time turns around $30$ minutes. We naturally have the same similarities
already given by the direct method.\newline

\noindent \textbf{Algorithm on the similarity by the theorem of
Glivenko-Cantelli} \newline

\noindent We randomly pick a number $NG=10000$ $k$-\textit{shinglings} from
both files and next we compute the similarity as we did in the case of the
direct method.\newline

\noindent We remark a considerable reduction of the time of the
determination of the similarity. The result is huge. The similarity indices
are got in less a minute. The similarity also turns around $50$ $\%$.

\bigskip

\noindent \textbf{Algorithm on the similarity by RUM}\newline

\noindent We randomly pick $N_{1}=10000$ \textit{k-shinglings} from of the
first file and $N_{2}$=10000 \textit{k-shinglings} from the second file. We
apply the RUM algorithm with a number of hashing $pp$ taking the values 5,
10, $15$, $20$. To guarantee the stability of results, the RUM method is
used fifty times (BB=50) and the average similarity has been reported out in
tables $(6.5)$, $(6.6)$, $(6.7)$ and $(6.8)$.\newline

\noindent Finally, we arrive at a tuning result : by using subsamples of the
two sets and by using the approximation method via the RUM algorithm, we get
an acceptable estimation of the similarity in a few number of seconds. But
since the results may be biased, performing the process a certain number of
times and reporting the average is better.\newline

\bigskip

\noindent We may study the variability of the results. If we proceed $BB=50$
times with $pp=20$ hash functions, the different obtained values for the
similarities present an empirical deviation of the order of $10\%$. This
means that the reported value is accurate at $2\%$.\newline

\noindent For the Gospels for example, we finally conclude that the true
estimation of the similarity is in an interval centered at the approximated
value given by the RUM method with magnitude $10\%$. This result, that is
achieved only in seconds, is very significant for large sets.

\bigskip

\noindent We may also appreciate the power of this algorithm that allows
estimation of the similarity of set around one hundred thousand (100.000)
characters in only 6 seconds.

\subsubsection{Comparison of Gospels}

\label{sss632}

\noindent From the tables $(6.5)$, $(6.6)$, $(6.7)$ and $(6.8)$, we notice
that the Gospels of Luke and Matthew have the greatest similarity around $70$
$\%$. From what we already said in Subsection \ref{ss61}, Luke and Matthew
have used the Gospel of Mark and in addition, are based on unknown source $Q 
$. Likewise the similarity between the Gospel of John and the others might
explained by the fact that the John Gospel is the last to be released in about year $%
100$ or year $110$ of our era. He might already be aware of the contents of
the other three gospels.\newline

\noindent We might hope to have a similarity around $90$ $\%$. But many
factors can influence on the outcomes. Actually, the Gospels are written by
four different persons. Each of them may use his own words. Besides, we used
translated versions. This latter fact can result in a significant decrease
of the true similarity. An other point concerns the fact that a limited
alphabet is used. This in turn is in favor of forming a structural part in
the similarity. For example, for the considered sizes, this part is around $%
30\%$. \newline

\noindent With the order of the sets sizes, we have the automatic and
stochastic similarity of order of $30\%$. Since the similarities turn around 
$50\%$ between the Gospels, we conclude that Gospels really have a
significant similarity. By taking account the remarks that have been made
above, we may expect that these similarities should be really much greater.
This is in favor of the hypothesis of the existence of a common source that
can be denamed as the source Q.

\subsubsection{Recommendations and perspectives}

\label{sss633}

\noindent To conclude we recommend these following steps in assessing
similarity :\newline

\begin{itemize}
\item[1.] Determine the automatic and stochastic part of the similarity, by
simulation studies by using formula (\ref{esp}).

\item[2.] Form the sets of \textit{k-shinglings} of the two studied sets.

\item[3.] Pick at random $n_{1}$ and $n_{2}$ \textit{k-shinglings } for the
two sets to study.

\item[4.] Apply the RUM algorithm.

\item[5.] Compare the finding similarity with the results of the point 
\textbf{(1)}.

\item[6.] Conclude on a significant similarity if the reached similarity, is
widely superior tothe stochastic similarity determined in (1). Otherwise the
similarity is not accepted.

\item[7.] Apply the RUM algorithm a number of times before doing definitive
conclusion.
\end{itemize}

\subsubsection{ Conclusion}

\label{sss634}

\noindent In this paper we described the main methods of determination of
the similarity. We empirically estimated the incompressible stochastic
similarity between two texts. We proposed a modification of the RU
algorithm, named RUM, and we applied on subsamples of the studied texts. The
combination of the Glivenko-Cantelli theorem and an empirical study of the
RUM algorithm, leads to the conclusion that the approximated similarity that
is given by this procedure, is a good estimation of the true similarity.
Since this approximated similarity is computed in seconds, the method showed
remarkable performance. Hence it is recommended for the study of similarity
for very large data sets.\newline

\noindent We applied our methods to the four Gospels. The obtained results
concern the study of Gospels themselves as well as the evaluation of
different methods of computation of the similarity. In conclusion, the
Gospels have indices of similarity at least $50\%$.\newline

\noindent In a coming paper we will concentrate on the theoretical
foundations of the RUM algorithm in the setting of Probability theory and
Statistics.\newline


\begin{thebibliography}{99}
\bibitem{AnandJeffrey} Anand R. and Jeffrey D. U. (2011). \textit{Mining of
Massive Datasets}. California.

\bibitem{GowerLegendre} J. C. Gower and P. Legendre. (1986). Metric and Euclidean
properties of dissimilarity coefficients. \textit{Journal of Classification}, 3, pp. 5-48.

\bibitem{EVan} http://fr.wikipedia.org/wiki/\'{E}vangiles date 07-04-2015 at
13:14

\bibitem{EVan1} http://www.info-bible.org/lsg/INDEX.html date 07-04-2013 at
12:21

\bibitem{SteinEissen06} Benno S. and Sven M. z. E.(2006). Near Similarity Search
and Plagiarism Analysis. In: \textit{  Spiliopoulou et al. (Eds.): From Data and Information Analysis to Knowledge Engineering
Selected Papers from the 29th Annual Conference of the German Classification Society (GfKl)
Magdeburg}, pp. 430-437, Springer.

\bibitem{ZezulaAmatoDohnal06} Pavel Z., Vlastislav D. and Giuseppe A. (2006). \textit{Similarity
Search The Metric Space Approach}, Springer.

\bibitem{Sung07} Sung-Hyuk C. (2007). Comprehensive Survey on Distance Similarity
Measures between Probability Density Functions. \textit{International jounal of mathematical models and methods in applied sciences}, 4(1), pp. 300-307.

\bibitem{DingXiaoYuMing03} Ding-Yun C. Xiao-Pei T. Yu-Te S. and Ming
O. ( 2003). On Visual Similarity Based 3D Model Retrieval, \textit{Eurographics 2003 / P. Brunet and D. Fellner
(Guest Editors)}, 22(3), pp. 223-232.

\bibitem{Formica05} Anna F. ( 2005). Ontology-based concept similarity in Formal
Concept Analysis, \textit{Information sciences}, pp. 2624-2641.

\bibitem{StrehlGhoshMooney00} Alexander S., Joydeep G., and Raymond M. (2000). Impact of
Similarity Measures on Web-page Clustering. \textit{American Association for Artificial Intelligence}, pp. 78712-1084.

\bibitem{BilenkoMooney03} Mikhail B. and Raymond J. M. (2003). Adaptive Duplicate
Detection Using Learnable String Similarity Measures.  In: \textit{ Proceedings of the Ninth ACM SIGKDD International Conference on Knowledge Discovery
and Data Mining(KDD)}, pp.39-48, Washington DC.

\bibitem{GionisIndykyMotwaniz} Aristides G., Piotr I. and Rajeev M. (1999).
Similarity Search in High Dimensions via Hashing. In: \textit{  Proceedings of the 25th VLDB Conference}, pp. 518-529 , Eds: Edinburgh, Scotland.

\bibitem{TheobaldSiddharth Paepcke08} Martin T., Jonathan S., and Andreas P. (2008). SpotSigs: robust and efficient near duplicate detection in large
web collections. \textit{ In:   31st Annual ACM SIGIR Conference}, Singapore.
\end{thebibliography}
\end{document}